\pdfoutput=1
\PassOptionsToPackage{unicode}{hyperref}
\PassOptionsToPackage{hyphens}{url}
\PassOptionsToPackage{dvipsnames,svgnames,x11names}{xcolor}
\documentclass[10pt,a4paper,onecolumn]{article}
\usepackage[T1]{fontenc}
\usepackage[utf8]{inputenc}
\usepackage{lmodern}
\usepackage[rgb,svgnames]{xcolor}
\usepackage{authblk,etoolbox}
\usepackage{titlesec}
\usepackage{calc}
\usepackage{tikz}
\usepackage{hyperref}
\hypersetup{%
    pdftitle={PLATON 7: An ultra-fast exoplanet atmospheric retrieval
code},
    pdfauthor={Michael Zhang, Tyler Fairnington, Yayaati Chachan,
Kimberly Paragas, Eliza Kempton, Heather Knutson, Jacob Bean},
    colorlinks=true,
    linkcolor=[rgb]{0.0, 0.5, 1.0},
    citecolor=Blue,
    urlcolor=[rgb]{0.0, 0.5, 1.0},
    breaklinks=true
}
\usepackage{orcidlink}
\usepackage{amssymb,amsmath}

\makeatletter
\@ifundefined{KOMAClassName}{
  \IfFileExists{parskip.sty}{%
    \usepackage{parskip}
  }{
    \setlength{\parindent}{0pt}
    \setlength{\parskip}{6pt plus 2pt minus 1pt}}
}{
  \KOMAoptions{parskip=half}}
\makeatother
\usepackage{longtable,booktabs,array}
\usepackage{calc} 
\usepackage{etoolbox}
\makeatletter
\patchcmd\longtable{\par}{\if@noskipsec\mbox{}\fi\par}{}{}
\makeatother
\IfFileExists{footnotehyper.sty}{\usepackage{footnotehyper}}{\usepackage{footnote}}
\makesavenoteenv{longtable}
\NewDocumentCommand\citeproctext{}{}
\NewDocumentCommand\citeproc{mm}{%
  \begingroup\def\citeproctext{#2}\cite{#1}\endgroup}
\makeatletter
 \let\@cite@ofmt\@firstofone
 \def\@biblabel#1{}
 \def\@cite#1#2{{#1\if@tempswa , #2\fi}}
\makeatother
\newlength{\cslhangindent}
\newlength{\csllabelwidth}
\newenvironment{CSLReferences}[2] 
 {\begin{list}{}{%
  \setlength{\itemindent}{0pt}
  \setlength{\leftmargin}{0pt}
  \setlength{\parsep}{0pt}
  \ifodd #1
   \setlength{\leftmargin}{\cslhangindent}
   \setlength{\itemindent}{-1\cslhangindent}
  \fi
  \setlength{\itemsep}{#2\baselineskip}}}
 {\end{list}}
\usepackage{calc}

\usepackage[margin=2.5cm]{geometry}

\titleformat{\section}
  {\normalfont\sffamily\Large\bfseries}
  {}{0pt}{}
\titleformat{\subsection}
  {\normalfont\sffamily\large\bfseries}
  {}{0pt}{}
\titleformat{\subsubsection}
  {\normalfont\sffamily\bfseries}
  {}{0pt}{}
\titleformat*{\paragraph}
  {\sffamily\normalsize}

\definecolor{c53baa1}{RGB}{83,186,161}
\definecolor{c202826}{RGB}{32,40,38}
\def \rorglobalscale {0.1}
\newcommand{\rorlogo}{%
\begin{tikzpicture}[y=1cm, x=1cm, yscale=\rorglobalscale,xscale=\rorglobalscale, every node/.append style={scale=\rorglobalscale}, inner sep=0pt, outer sep=0pt]
  \begin{scope}[even odd rule,line join=round,miter limit=2.0,shift={(-0.025, 0.0216)}]
    \path[fill=c53baa1,nonzero rule,line join=round,miter limit=2.0] (1.8164, 3.012) -- (1.4954, 2.5204) -- (1.1742, 3.012) -- (1.8164, 3.012) -- cycle;
    \path[fill=c53baa1,nonzero rule,line join=round,miter limit=2.0] (3.1594, 3.012) -- (2.8385, 2.5204) -- (2.5172, 3.012) -- (3.1594, 3.012) -- cycle;
    \path[fill=c53baa1,nonzero rule,line join=round,miter limit=2.0] (1.1742, 0.0669) -- (1.4954, 0.5588) -- (1.8164, 0.0669) -- (1.1742, 0.0669) -- cycle;
    \path[fill=c53baa1,nonzero rule,line join=round,miter limit=2.0] (2.5172, 0.0669) -- (2.8385, 0.5588) -- (3.1594, 0.0669) -- (2.5172, 0.0669) -- cycle;
    \path[fill=c202826,nonzero rule,line join=round,miter limit=2.0] (3.8505, 1.4364).. controls (3.9643, 1.4576) and (4.0508, 1.5081) .. (4.1098, 1.5878).. controls (4.169, 1.6674) and (4.1984, 1.7642) .. (4.1984, 1.8777).. controls (4.1984, 1.9719) and (4.182, 2.0503) .. (4.1495, 2.1132).. controls (4.1169, 2.1762) and (4.0727, 2.2262) .. (4.0174, 2.2635).. controls (3.9621, 2.3006) and (3.8976, 2.3273) .. (3.824, 2.3432).. controls (3.7505, 2.359) and (3.6727, 2.367) .. (3.5909, 2.367) -- (2.9676, 2.367) -- (2.9676, 1.8688).. controls (2.9625, 1.8833) and (2.9572, 1.8976) .. (2.9514, 1.9119).. controls (2.9083, 2.0164) and (2.848, 2.1056) .. (2.7705, 2.1791).. controls (2.6929, 2.2527) and (2.6014, 2.3093) .. (2.495, 2.3487).. controls (2.3889, 2.3881) and (2.2728, 2.408) .. (2.1468, 2.408).. controls (2.0209, 2.408) and (1.905, 2.3881) .. (1.7986, 2.3487).. controls (1.6925, 2.3093) and (1.6007, 2.2527) .. (1.5232, 2.1791).. controls (1.4539, 2.1132) and (1.3983, 2.0346) .. (1.3565, 1.9436).. controls (1.3504, 2.009) and (1.3351, 2.0656) .. (1.3105, 2.1132).. controls (1.2779, 2.1762) and (1.2338, 2.2262) .. (1.1785, 2.2635).. controls (1.1232, 2.3006) and (1.0586, 2.3273) .. (0.985, 2.3432).. controls (0.9115, 2.359) and (0.8337, 2.367) .. (0.7519, 2.367) -- (0.1289, 2.367) -- (0.1289, 0.7562) -- (0.4837, 0.7562) -- (0.4837, 1.4002) -- (0.6588, 1.4002) -- (0.9956, 0.7562) -- (1.4211, 0.7562) -- (1.0118, 1.4364).. controls (1.1255, 1.4576) and (1.2121, 1.5081) .. (1.2711, 1.5878).. controls (1.2737, 1.5915) and (1.2761, 1.5954) .. (1.2787, 1.5991).. controls (1.2782, 1.5867) and (1.2779, 1.5743) .. (1.2779, 1.5616).. controls (1.2779, 1.4327) and (1.2996, 1.3158) .. (1.3428, 1.2113).. controls (1.3859, 1.1068) and (1.4462, 1.0176) .. (1.5237, 0.944).. controls (1.601, 0.8705) and (1.6928, 0.8139) .. (1.7992, 0.7744).. controls (1.9053, 0.735) and (2.0214, 0.7152) .. (2.1474, 0.7152).. controls (2.2733, 0.7152) and (2.3892, 0.735) .. (2.4956, 0.7744).. controls (2.6016, 0.8139) and (2.6935, 0.8705) .. (2.771, 0.944).. controls (2.8482, 1.0176) and (2.9086, 1.1068) .. (2.952, 1.2113).. controls (2.9578, 1.2253) and (2.9631, 1.2398) .. (2.9681, 1.2544) -- (2.9681, 0.7562) -- (3.3229, 0.7562) -- (3.3229, 1.4002) -- (3.4981, 1.4002) -- (3.8349, 0.7562) -- (4.2603, 0.7562) -- (3.8505, 1.4364) -- cycle(0.9628, 1.7777).. controls (0.9438, 1.7534) and (0.92, 1.7357) .. (0.8911, 1.7243).. controls (0.8623, 1.7129) and (0.83, 1.706) .. (0.7945, 1.7039).. controls (0.7588, 1.7015) and (0.7252, 1.7005) .. (0.6932, 1.7005) -- (0.4839, 1.7005) -- (0.4839, 2.0667) -- (0.716, 2.0667).. controls (0.7477, 2.0667) and (0.7805, 2.0643) .. (0.8139, 2.0598).. controls (0.8472, 2.0553) and (0.8768, 2.0466) .. (0.9025, 2.0336).. controls (0.9282, 2.0206) and (0.9496, 2.0021) .. (0.9663, 1.9778).. controls (0.9829, 1.9534) and (0.9914, 1.9209) .. (0.9914, 1.8799).. controls (0.9914, 1.8362) and (0.9819, 1.8021) .. (0.9628, 1.7777) -- cycle(2.6125, 1.3533).. controls (2.5889, 1.2904) and (2.5553, 1.2359) .. (2.5112, 1.1896).. controls (2.4672, 1.1433) and (2.4146, 1.1073) .. (2.3529, 1.0814).. controls (2.2916, 1.0554) and (2.2228, 1.0427) .. (2.1471, 1.0427).. controls (2.0712, 1.0427) and (2.0026, 1.0557) .. (1.9412, 1.0814).. controls (1.8799, 1.107) and (1.8272, 1.1433) .. (1.783, 1.1896).. controls (1.7391, 1.2359) and (1.7052, 1.2904) .. (1.6817, 1.3533).. controls (1.6581, 1.4163) and (1.6465, 1.4856) .. (1.6465, 1.5616).. controls (1.6465, 1.6359) and (1.6581, 1.705) .. (1.6817, 1.7687).. controls (1.7052, 1.8325) and (1.7388, 1.8873) .. (1.783, 1.9336).. controls (1.8269, 1.9799) and (1.8796, 2.0159) .. (1.9412, 2.0418).. controls (2.0026, 2.0675) and (2.0712, 2.0804) .. (2.1471, 2.0804).. controls (2.223, 2.0804) and (2.2916, 2.0675) .. (2.3529, 2.0418).. controls (2.4143, 2.0161) and (2.467, 1.9799) .. (2.5112, 1.9336).. controls (2.5551, 1.8873) and (2.5889, 1.8322) .. (2.6125, 1.7687).. controls (2.636, 1.705) and (2.6477, 1.6359) .. (2.6477, 1.5616).. controls (2.6477, 1.4856) and (2.636, 1.4163) .. (2.6125, 1.3533) -- cycle(3.8015, 1.7777).. controls (3.7825, 1.7534) and (3.7587, 1.7357) .. (3.7298, 1.7243).. controls (3.701, 1.7129) and (3.6687, 1.706) .. (3.6333, 1.7039).. controls (3.5975, 1.7015) and (3.5639, 1.7005) .. (3.5319, 1.7005) -- (3.3226, 1.7005) -- (3.3226, 2.0667) -- (3.5547, 2.0667).. controls (3.5864, 2.0667) and (3.6192, 2.0643) .. (3.6526, 2.0598).. controls (3.6859, 2.0553) and (3.7155, 2.0466) .. (3.7412, 2.0336).. controls (3.7669, 2.0206) and (3.7883, 2.0021) .. (3.805, 1.9778).. controls (3.8216, 1.9534) and (3.8301, 1.9209) .. (3.8301, 1.8799).. controls (3.8301, 1.8362) and (3.8206, 1.8021) .. (3.8015, 1.7777) -- cycle;
  \end{scope}
\end{tikzpicture}
}

\makeatletter
\patchcmd{\@maketitle}{center}{flushleft}{}{}
\patchcmd{\@maketitle}{center}{flushleft}{}{}
\patchcmd{\@maketitle}{\LARGE}{\LARGE\sffamily}{}{}
\def\maketitle{{%
  
  \AB@maketitle}}
\renewcommand\AB@affilsepx{ \protect\Affilfont}
\renewcommand\AB@affilnote[1]{{\bfseries #1}\hspace{3pt}}
\renewcommand{\affil}[2][]%
   {\newaffiltrue\let\AB@blk@and\AB@pand
      \if\relax#1\relax\def\AB@note{\AB@thenote}\else\def\AB@note{#1}%
        \setcounter{Maxaffil}{0}\fi
        \begingroup
        \let\href=\href@Orig
        \let\protect\@unexpandable@protect
        \def\thanks{\protect\thanks}\def\footnote{\protect\footnote}%
        \@temptokena=\expandafter{\AB@authors}%
        {\def\\{\protect\\\protect\Affilfont}\xdef\AB@temp{#2}}%
         \xdef\AB@authors{\the\@temptokena\AB@las\AB@au@str
         \protect\\[\affilsep]\protect\Affilfont\AB@temp}%
         \gdef\AB@las{}\gdef\AB@au@str{}%
        {\def\\{, \ignorespaces}\xdef\AB@temp{#2}}%
        \@temptokena=\expandafter{\AB@affillist}%
        \xdef\AB@affillist{\the\@temptokena \AB@affilsep
          \AB@affilnote{\AB@note}\protect\Affilfont\AB@temp}%
      \endgroup
       \let\AB@affilsep\AB@affilsepx
}
\makeatother

\renewcommand\Affilfont{\sffamily\small\mdseries}
\IfFileExists{upquote.sty}{\usepackage{upquote}}{}
\IfFileExists{microtype.sty}{
  \usepackage[]{microtype}
  \UseMicrotypeSet[protrusion]{basicmath} 
}{}

\PassOptionsToPackage{usenames,dvipsnames}{color} 
\title{PLATON 7: An ultra-fast exoplanet atmospheric retrieval code}

\author[1%
\ensuremath\mathparagraph]{Michael Zhang%
  \,\orcidlink{0000-0002-0659-1783}\,%
}
\author[1%
]{Tyler Fairnington%
}
\author[2%
]{Yayaati Chachan%
}
\author[3%
]{Kimberly Paragas%
}
\author[1%
]{Eliza Kempton%
}
\author[3%
]{Heather Knutson%
}
\author[1%
]{Jacob Bean%
}

\affil[1]{University of Chicago%
  \,\protect\href{https://ror.org/024mw5h28}{\protect\rorlogo}\,%
}
\affil[2]{UC Santa Cruz%
}
\affil[3]{Caltech%
}
\affil[$\mathparagraph$]{Corresponding author}
\date{\vspace{-2.5ex}}

\begin{document}
\maketitle

\section{Summary}\label{summary}

The atmospheres of hundreds of exoplanets have been observed with
transmission or emission spectroscopy. To derive composition,
temperature, cloudtop pressures, and other properties from these
observations, the most common tool is a retrieval: a parameterized
atmospheric model, coupled with a Bayesian exploration of parameter
space using nested sampling or MCMC. With the launch of JWST and
proliferation of high-resolution spectrographs on ground telescopes, the
increasing precision and volume of exoplanet atmospheric spectra require
retrieval codes that are fast even at high spectral resolution.

\section{Statement of need}\label{statement-of-need}

PLATON is an open source atmospheric retrieval Python code first
released in 2018 (\citeproc{ref-Zhang:2019}{Zhang et al., 2019}) that
was greatly expanded in Zhang et al. (\citeproc{ref-Zhang:2020}{2020})
and Zhang et al. (\citeproc{ref-Zhang:2025}{2025}). The forward model
can generate transmission and emission of both rocky and gaseous
exoplanets. The retrieval supports both equilibrium chemistry and free
retrieval using either nested sampling or MCMC.

PLATON 7 is a complete rewrite of PLATON 6 using Claude Fable 5. The
biggest changes are that PLATON now uses JAX
(\citeproc{ref-Bradbury:2018}{Bradbury et al., 2018}) for all intensive
calculations, and these calculations are now performed in single
precision instead of double precision. These two changes speed up the
code by a factor of \textasciitilde60 with negligible decrease in
accuracy, enabling a 0.8–5.3 um transmission spectrum to be computed at
R=20k in 0.35 ms on a RTX 5090. To prevent underflows or overflows in
intermediate calculations, we instructed Claude to rearrange operations
and rescale values (notably opacities and polarizabilities) to be near
unity.

An unrelated change which contributes to the vast speedup is our switch
to nautilus (\citeproc{ref-Lange:2023}{Lange, 2023}) as the default
nested sampling package. Unlike classic nested sampling packages such as
MultiNest (\citeproc{ref-Feroz:2009}{Feroz et al., 2009}) or dynesty
(\citeproc{ref-Speagle:2020}{Speagle, 2020}), nautilus trains small
neural networks to draw the live set boundary instead of fitting
hyper-ellipsoids or other geometric objects. In our experience, nautilus
enables far more efficient sampling at high dimensionality, in addition
to more accurate identification of small (low prior volume) but high
likelihood modes.

In addition to the large speedup, we add a variety of features, as
detailed in our documentation and examples. For example, it is now easy
for users to run a forward model with custom temperature and abundance
profiles. We add support for two-limb retrievals because JWST
observations have revealed the importance of limb asymmetry in gas giant
transit spectra (\citeproc{ref-Fairnington:2026}{Fairnington et al.,
2026}). We also add support for non-isothermal limbs in transit spectra,
again motivated by the increasing precision of data.

\section{State of the field}\label{state-of-the-field}

As of March 2023, at least 50 retrieval codes existed in the exoplanet
literature, of which 26 were open source
(\citeproc{ref-MacDonaldBatalha:2023}{MacDonald \& Batalha, 2023}).
However, highly optimized GPU retrieval codes are much rarer. We are
only aware of one: BeAR (\citeproc{ref-Wang:2026}{Wang et al., 2026}), a
successor to HELIOS-r2 (\citeproc{ref-Kitzmann:2020}{Kitzmann et al.,
2020}), written in C++ and CUDA with a Python wrapper.

Because the main motivation of our PLATON 7 rewrite was speed, we
benchmarked it against 4 widely used retrieval codes: PLATON 6
(\citeproc{ref-Zhang:2025}{Zhang et al., 2025}), POSEIDON 1.4
(\citeproc{ref-MacDonald:2023}{MacDonald, 2023};
\citeproc{ref-MacDonald:2017}{MacDonald \& Madhusudhan, 2017}), TauREx
3.3 (\citeproc{ref-AlRefaie:2021}{Al-Refaie et al., 2021}), and BeAR 2
(\citeproc{ref-Wang:2026}{Wang et al., 2026}). We ran free retrievals
with the same parameterization on the transit spectrum of TOI 6894b
(\citeproc{ref-Zhang:2026}{Zhang et al., 2026}), making sure that the
spectral resolution and number of atmospheric layers are also identical
across all codes. All benchmarks were run on a vast.ai virtual instance
with a AMD Ryzen 9 7950X3D 16-core CPU, 45 GB of RAM, a RTX 5090 GPU,
and CUDA 13.0. The CPU codes were run with 16 cores for POSEIDON and
TauREx, and 10 cores (the maximum that the memory allows for) for PLATON
7 with GPU use disabled.

\begin{longtable}[]{@{}lll@{}}
\toprule\noalign{}
Code & Time (ms) & GPU? \\
\midrule\noalign{}
\endhead
\bottomrule\noalign{}
\endlastfoot
PLATON 7 & 0.35 & Yes \\
BeAR & 0.61 & Yes \\
PLATON 6 & 20 & Yes \\
TauREx-CUDA & 17 & Yes \\
PLATON 7 & 14 & No \\
POSEIDON & 45 & No \\
TauREx & 132 & No \\
\end{longtable}

The table above reports the total wall clock time divided by the total
number of forward models, after discarding the initialization and 5000
forward models’ worth of burn-in. PLATON 7 is substantially faster than
other codes we tested, beating BeAR by 1.75x, PLATON 6 by 57x, and
CPU-based codes by around 2-3 orders of magnitude. When not allowed to
use the GPU, PLATON 7 transparently falls back to the CPU and remains
substantially faster than other CPU codes.

The speedup enabled by nautilus is hard to quantify because it is highly
dependent on the specific retrieval problem, but we have seen speedups
over MultiNest of more than 80x for retrievals with many parameters.
This is especially the case for retrievals where each limb is fit
separately with a non-isothermal T/P profile.

\section{Software design}\label{software-design}

PLATON was initially designed to be fast, open source, easy to install
and use, and easy to understand. To this end, we implemented it in pure
Python and kept the functionality spartan, enabling us to keep the
number of lines to a minimum. The idea was that users could easily hack
the code to implement the functionality they wanted. This idea was
successful, and researchers from other groups implemented free
retrievals, patchy clouds, and rocky surface emission long before they
were included in the official PLATON.

With the advent of powerful AI coding assistants, we de-emphasized the
last design goal– ease of human understanding–in favor of the first, raw
speed. Our experience is that frontier LLMs like Fable are fully capable
of understanding the code and making changes to it. As AI becomes more
powerful, we expect only a vanishingly small portion of PLATON users
will read the code themselves, making human readability much less
important. That said, we have not sacrificed this goal completely:
PLATON 7 is still in pure Python, retains largely the same structure as
PLATON 6, and implements the same core algorithms.

PLATON 7 remains easy to install and use, even without AI assistance.
Its user-facing interface is nearly identical to that of its
predecessor. We have added more example scripts and simplified the ones
that already exist in order to flatten the learning curve even further
for humans and AIs alike.

\section{Research impact statement}\label{research-impact-statement}

PLATON is one of the oldest and most widely used open source retrieval
codes.\\
It has been used by more than 80 papers to investigate everything from
sulfur biosignatures on sub-Neptunes (\citeproc{ref-Tsai:2024}{Tsai et
al., 2024}) to aerosols on warm sub-Saturns
(\citeproc{ref-Estrela:2021}{Estrela et al., 2021}). The majority of
these papers do not involve the main developer. Users span institutions
and career stages, and include undergraduate students in exoplanet
courses.

PLATON 7 specifically has been used to analyze the impact of limb
asymmetry on the transit spectrum of WASP-94b
(\citeproc{ref-Fairnington:2026}{Fairnington et al., 2026}). We expect
much wider adoption once the code is officially released, as the 57x
speedup over its precedessor enables use cases that would otherwise be
prohibitively expensive.

\section{Our vibecoding experience}\label{our-vibecoding-experience}

We expect that the majority of new software will be written by AI. Since
PLATON 7 is the first AI-written retrieval code that we are aware of, we
share our vibecoding experiences in the hopes that it will be useful for
the community.

Many years ago, the lead developer MZ aspired to make PLATON
auto-differentiable by basing it on JAX instead of cupy
(\citeproc{ref-Okuta:2017}{Okuta et al., 2017}), enabling the use of
sample-efficient Hamiltonian Monte Carlo methods. However, he dropped
the project after realized that JAX is not a drop-in replacement for
cupy, and that adopting JAX would mean rewriting the entire codebase.

In April 2026, we asked Claude Code with the Opus 4.6 model to implement
HMC with JAX-computed gradients. Opus first misunderstood our intent and
computed numerical gradients instead of rewriting the code to support
autodiff. After we clarified, it successfully rewrote the transit
spectrum code, but encountered numerical difficulties with the opaque
cloud deck, which we circumvented by adopting a continuous cloud
prescription. After more debugging, we realized that HMC struggles with
the posterior shapes typical in free retrievals. We therefore disabled
autodiff and switched back to nested sampling, enabling a significant
speedup over the old PLATON.

We then realized that a further speedup is possible by using fp32
instead of fp64 precision. Opus made the change and converted all of the
arrays to fp32, but reported no speedup. Realizing this made no sense,
we examined the AI’s progress reports and realized it inexplicably set
jax\_enable\_x64 to True. We set the variable to False and the code
instantly became many times faster, evaluating every forward model in
only 1.5 ms. Opus declared that the code was unusable because fp32 is
insufficient precision, but we told it to work around overflows and
underflows by reordering operations and rescaling variables to order
unity. The code then produced correct results for our test case.
However, over the following months, we found other bugs stemming from
the lower precision, which we fixed one by one.

In September 2026, we started from scratch and prompted Fable 5 to
rewrite all of PLATON–not just a small part of the transit spectrum
code–to use JAX. We told it from the outset to use single precision, set
jax\_enable\_x64 to False, and prevent underflows or overflows by
rescaling variables to order unity and/or reordering operations.
Afterwards, we told it to optimize the code while ensuring that the unit
tests still pass. After it was finished, we asked it to ask Codex
(GPT-5.6 Sol) to review its changes and fix any important issues. We
also triggered /code-review to have Claude review its own changes, then
prompted it to fix any important issues.

This time, Claude was able to perform the entire refactoring and
optimization process with no apparent issues and minimal human guidance.
We have exercised the code extensively and could not find any new bugs.
We attribute this to a combination of the much stronger model (Fable 5
vs.~Opus 4.6), more specific initial instructions, and extensive code
review by both GPT-5.6 Sol and Fable. Our experience is that frontier
LLMs already outperform nearly all, if not all, astronomers at writing
correct and fast code.

\section{AI usage disclosure}\label{ai-usage-disclosure}

Fable 5 with the Claude Code harness did all of the code rewrite.
GPT-5.6 Sol with the Codex harness reviewed the code for correctness.
More features were subsequently added by Opus 5.5.

No AI was used in the writing of this manuscript, except for finding and
adding references.

\section{Acknowledgements}\label{acknowledgements}

We acknowledge contributions from Arjun Savel and Vatsal Panwar during
the development of PLATON.

MZ acknowledges funding from the Heising-Simons Foundation for his 51
Pegasi b postdoctoral fellowship.

\section*{References}\label{references}
\addcontentsline{toc}{section}{References}

\protect\phantomsection\label{refs}

\end{document}